\documentclass[preprint,authoryear,12pt]{elsarticle}

\usepackage{graphicx}
\usepackage{a4wide}
\usepackage{booktabs}
\usepackage{natbib} 
\usepackage{amsfonts}
\usepackage{amsmath}
\usepackage{amssymb}
 \usepackage{amsthm}
 \usepackage{epsfig}

\journal{XYZ}

\begin{document}

\begin{frontmatter}

\title{Long-term memory in electricity prices: Czech market evidence}

\author[utia,ies]{Ladislav Kristoufek} \ead{kristouf@utia.cas.cz}
\author[ies]{Petra Lunackova} \ead{lunackova.petra@gmail.com }

\address[utia]{Institute of Information Theory and Automation, Academy of Sciences of the Czech Republic, Pod Vodarenskou Vezi 4, 182 08, Prague, Czech Republic, EU} 
\address[ies]{Institute of Economic Studies, Faculty of Social Sciences, Charles University in Prague, Opletalova 26, 110 00, Prague, Czech Republic, EU}

\begin{abstract}
We analyze long-term memory properties of hourly prices of electricity in the Czech Republic between 2009 and 2012. As the dynamics of the electricity prices is dominated by cycles -- mainly intraday and daily -- we opt for the detrended fluctuation analysis, which is well suited for such specific series. We find that the electricity prices are non-stationary but strongly mean-reverting which distinguishes them from other financial assets which are usually characterized as unit root series. Such description is attributed to specific features of electricity prices, mainly to non-storability. Additionally, we argue that the rapid mean-reversion is due to the principles of electricity spot prices. These properties are shown to be stable across all studied years.
\end{abstract}


\end{frontmatter}

\newpage

\section{Introduction}

Electricity is a flow commodity with unique characteristics that influence the way it is traded and thus the behavior of spot and futures prices in the market. Electricity cannot be effectively stored (with minor exception of pumped-storage hydro power plants that are scarce) so that the adjustment of demand and supply must be instantaneous. Electricity consumption reflects human behavior and temporal patterns of human life with daily and weekly routines which is reflected in the daily pattern with single or double peak structure and weekly patterns \citep{simonsen2004structure}. On higher scales, the seasonal fluctuations are mainly caused by weather, in particular temperature and number of hours of daylight \citep{lucia2002electricity}. The seasonal patterns are strongly geographically dependent -- in northern countries, the highest consumption is usually observed during winter months due to heating, and in southern countries, air-conditioning increases the consumption during summer \citep{zachmann2008electricity}. 

Electricity prices on the spot market are very sensitive to temperatures and especially to sudden unexpected weather changes, which are expected up to a certain point. The weather forecast is never perfect which causes the spot prices to be much more volatile than other financial assets \citep{asbury1975weather}. Moreover, electricity supply side is also weather dependent, which is evidently more valid for the renewable sources of energy such as wind turbines and photovoltaic power plants \citep{von2010large}. 

Demand for electricity is highly inelastic. In short run, it is absolutely inelastic so that the price is determined by the supply curve (merit order curve, marginal cost curve) completely. The curve resembles upward sloping stairway, each step approximately represents a different type of a power plant and thus a different level of marginal costs. The price on the market rises until it reaches the marginal costs for a MWh of the power plant of the next level, after that the supply rises. This is why the merit order curve is not smooth. In order to produce an additional MWh,  more expensive power sources (plants) are activated and as the supply is increasing, the price increases as well \citep{geman2006understanding,sensfuss2008merit}.

High (or excess) volatility is another typical feature of electricity prices and it is mainly due to the non-storability of electricity itself. There are no reserves that could be used in case of sudden increase in demand or weather change \citep{janczura2013identifying}. The prices are not only volatile, the volatility has also a tendency to cluster. Apart from the clustering, volatility is also characteristic by an inverse leverage effect -- positive shocks increase price volatility more than the negative ones \citep{knittel2005empirical}. In addition, the electricity prices tend to ``jump'' very frequently. These jumps are usually referred to as ``spikes'' and these are typical by a sharp increase followed by a slower decrease causing pronounced asymmetry. Due to the properties described above, the electricity prices are often treated as non-stationary.

Unlike other financial time series, specifically prices of various assets, the electricity prices are mean reverting \citep{Simonsen2003,Weron2000}. According to \cite{barlow2002diffusion}, estimates of time for mean reversion are from two to six days. \cite{geman2005energy} states that with constant or slightly increasing demand, supply side is able to adjust the pattern so that the prices remain close to their mean value. 

The electricity prices are also influenced by factors that are unthinkable for other ``typical'' financial assets -- technical constraints. Power plant which is out of order due to either technical problems or regular maintenance can influence the price because the number of power plants is small and limited. Electricity can be easily and quickly transported but transmission lines have capacity constraints which must not be exceeded. That is the main reason why electricity prices differ in neighboring areas but it can also cause high levels of volatility due to potential instability of the whole system \citep{borenstein1997competitive}.

Last but not least, the electricity demand and thus also the prices depend on business cycle, economic activity or growth. Electricity consumption and economic growth are bounded; different studies suggest different direction of the causality, from electricity consumption to GDP, vice versa or both \citep{soytas2003energy,lee2005energy,squalli2007electricity,ciarreta2010economic}.

In our study, we focus on various properties of the electricity prices in the Czech Republic with a special attention put on long-term memory of the spot prices. The electricity market in the Czech Republic has been fully deregulated since 2006. The network of power plants consists of the less expensive hydro and nuclear power plants, the more expensive hard coal and gas power plants, with lignite plants somewhere in between \citep{sensfuss2008merit}. Small increase in demand can thus put into function considerably more expensive power plants and the occurrence of spikes is potentially high.

OTE (Czech electricity and gas market operator, established in 2001) organizes the day-ahead spot electricity market since 2002 which has been coupled through implicit auctions with the organized day-ahead electricity market in the Slovak Republic since 2009 and with the day-ahead electricity market in Hungary since 2012.  This market has a form of the daily auction, the traded period is 1 hour and minimum tradable volume 1 MWh, the trading currency is EUR. The market closes always the day before at 11AM. Volume of electricity registered in the OTE system in 2012 for day-ahead market was 10,971 GWh for sale and 10,562 GWh for purchase\footnote{Details are available at https://www.ote-cr.cz/statistics.}.

In the paper, we describe temporal patterns, distributional properties and mainly the correlation structure with a special attention on long-term memory of the prices. To do so, we utilize the detrended fluctuation analysis, which is well suited for time series with such a complicated structure as the electricity spot prices. We show that the prices are non-stationary, strongly persistent but they remain strongly mean reverting which well distinguishes them from other financial prices such as stocks and exchange rates which follow random walk pattern \citep{Cont2001}. The paper is structured as follows. Section 2 focuses on recent studies on long-term memory properties of electricity prices to which we mostly contribute. Section 3 presents the data, subsequent Section 4 describes the methodology. Section 5 discusses the results and Section 6 concludes.

\section{Brief literature review}

Correlations and memory characteristics of electricity prices have been a frequent object of interest of many studies in recent years. \cite{Weron2000} analyze California Power Exchange (CalPX) and Swiss Electricity hourly prices using the rescaled range analysis finding mean-reverting characteristics. The analysis is then broadened by \cite{Weron2002} who studies four electricity markets (CalPX, Nord Pool, Entergy and UK spot) with three different methods (rescaled range analysis, detrended fluctuation analysis and periodogram methods) and confirms that returns of the electricity prices are anti-persistent. \cite{Simonsen2003} analyzes the Nord Pool prices using multi-scale wavelet approach and compares it with the standard rescaled range analysis to show that the returns of electricity prices are weakly anti-persistent. The author stresses that a choice of an appropriate technique for the long-term memory estimation is crucial. \cite{Park2006} examine 11 US electricity markets using the vector autoregression methodology but importantly finds several price series to be stationary which is well against a standard understanding of prices of financial assets which are typically a unit root series and thus strongly non-stationary.

\cite{Koopman2007} develop an adjusted fractionally integrated autoregressive moving average model with generalized autoregressive conditional heteroskedasticity (ARFIMA-GARCH), which is able to capture day-of-the-week patterns and extreme price movements, specifically for the electricity prices and on three European markets (German EEX, French Powernext and Dutch APX), they show that the weekly patterns are indeed crucial in the daily prices analysis. \cite{Norouzzadeh2007} study long-term memory and multifractality of the Spanish spot market finding persistent yet strongly mean-reverting prices. \cite{Erzgraber2008} focus on long-term memory in Nord Pool markets and find the returns to be weakly (compared to the previous study) anti-persistent. They also find that the strength of memory depends on the daytime of the measurement, i.e. the prices are not only correlated from hour to hour but also in the same hour from day to day. Moreover, they show that the memory parameter varies strongly in time. \cite{Uritskaya2008} examine the electricity prices of Alberta and Mid-C electricity prices using the detrended fluctuation analysis and spectral exponents finding that both the Alberta and Mid-C prices are persistent and mean-reverting. However, the former remains stationary whereas the latter does not.

\cite{Malo2009} combines various properties of electricity prices and utilizes Markov-switching multifractal model with conditional copulas to construct a model for risk minimization of the Nord Pool markets. Comparing various methods of long-term memory estimation, the author finds anti-persistent returns of electricity prices. Utilizing various copula specifications, conditional value at risk is also discussed in detail.

\cite{Alvarez-Ramirez2010} analyze Ontario and Alberta electricity markets with the detrended fluctuation analysis and the Allan factor model to show that the long-term memory properties of both prices and demand strongly vary in time. \cite{Haugom2011} model Nord Pool electricity prices using long-term memory mimicking heterogenous autoregressive model with realized variance (HAR-RV) and show that incorporating the strongly persistent realized variance improves the predicting power of the model. And \cite{Rypdal2013} model the Nord Pool data using the Multifractal random walk model adjusted for mean-reversion and volatility persistence to capture the most important characteristics of the electricity prices. Using the model, the authors show that the electricity prices characteristics are very different from the ones of the stock market prices. In our analysis, we apply the detrended fluctuation analysis on hourly prices of the Czech electricity. Specifically, we utilize its ability to separate cycles and seasonalities from the long-term memory.

\section{Methodology}

\subsection{Long-term memory}

Long-term memory is traditionally connected to slowly decaying auto-correlation functions. For the auto-correlation function $\rho(k)$ with a lag $k$, the decay is described as asymptotically hyperbolic so that $\rho(k)\propto k^{2H-2}$ where $k \rightarrow +\infty$. The auto-correlation function thus follows and asymptotic power law. A characteristic parameter of the long-term memory is the Hurst exponent $H$ which ranges between 0 and 1 for stationary processes. The breaking value of 0.5 is connected to a short-term correlated process (usually characteristic by exponential or more rapid decay of the auto-correlation function). For $H>0.5$, the underlying process is positively correlated and locally trending, and it is traditionally labeled as a persistent process. For $H<0.5$, the process is anti-persistent and it switches its direction more frequently than a random process would. The anti-persistent processes are also labelled as mean-reverting process as they return to the long-term mean very quickly \citep{Beran1994,Samorodnitsky2006}.

For non-stationary processes, the definition of long-term memory via the auto-correlation function suffers as the process has infinite variance and the correlations do not exist. For this matter, but also in a general case, a spectrum-based definition is used. Assuming that the spectrum or pseudo-spectrum of an underlying process exists near to the origin, i.e. $f(\lambda)$ exists for $\lambda\rightarrow 0+$, we define long-term memory via a power-law at origin spectrum, i.e. $f(\lambda) \propto \lambda^{1-2H}$ for $\lambda\rightarrow 0+$. For persistent processes, $f(\lambda)$ diverges at the origin whereas for the anti-persistent processes, it collapses to zero \citep{Samorodnitsky2006}.

Historically, there have been two major streams of the Hurst exponent estimators -- time domain estimators and frequency domain estimators. The time domain estimators are based on the auto-correlation definition of long-term memory and its implications to a scaling of variance of partial sums. To name the most frequently used ones, we have rescaled range analysis \citep{Hurst1951,Mandelbrot1968,Mandelbrot1968a}, detrended fluctuation analysis \citep{Peng1993,Peng1994,Kantelhardt2002}, generalized Hurst exponent approach \citep{Alvarez-Ramirez2002,DiMatteo2003,DiMatteo2007} and detrending moving average \citep{Alessio2002}. The frequency domain estimators are based on the spectrum definition and among the most popular ones are GPH estimator \citep{Geweke1983}, average periodogram estimator \citep{Robinson1994}, log-periodogram estimator \citep{Beran1994,Robinson1995} and local Whittle estimator \citep{Kunsch1987,Robinson1995a}. Due to very specific statistical properties of the electricity prices that have been mentioned in the previous sections and are also discussed in the following section, we opt for the detrended fluctuation analysis which has desirable properties for such type of analysis.

\subsection{Detrended fluctuation analysis}

Detrended fluctuation analysis (DFA) of \cite{Peng1993,Peng1994} is a special case of the multifractal detrended fluctuation analysis (MF-DFA) introduced by \cite{Kantelhardt2002}. For better understanding of the procedure, we present the more general MF-DFA as an initial step.

Let's have a time series $\{x_t\}$ with $t=1,\ldots,T$ where $T$ is a finite time series length. The profile $X(t)$ is constructed as 
\begin{equation}
X(t)=\sum_{i=1}^{t}{(x_i-\bar{x})}
\end{equation}
where $\bar{x}=\frac{1}{T}\sum_{t=1}^{T}x_t$ is a time series average. The profile is then divided into $T_s\equiv \lfloor T/s \rfloor$ non-overlapping windows with length $s$ (scale) where $\lfloor \rfloor$ is a lower integer part operator. As $T_s$ is not necessarily equal to $T/s$, part of the time series is left at the end of the series. In order not to lose the information of this segment, the profile is also divided from the opposite end and both sets of blocks of length $s$ are further utilized (we thus get $2T_s$ windows of length $s$).

In each of these $2T_s$ segments, we calculate the mean squared deviation from the trend of the series in this particular window. This means that for the $k$th window, the mean squared deviation $F^2(k,s)$ is obtained as
\begin{equation}
F^2(k,s)=\frac{1}{s}\sum_{i=1}^{s}{(X(s[k-1]+i)-\widehat{X_k(i)})^2}
\end{equation}
where $\widehat{X_k(i)}$ is a polynomial fit of time trend at position $i$ in window $k$. In our application, we utilize the linear fit obtained via standard ordinary least squares regression which is standard for the DFA and MF-DFA procedures \citep{Hu2001,Grech2005,Kantelhardt2009,Kristoufek2010}. This is applied for windows $k=1,\ldots,T_s$, and then for windows $k=T_s+1,\ldots,2T_s$, we obtain
\begin{equation}
F^2(k,s)=\frac{1}{s}\sum_{i=1}^{s}{(X(T-s[k-T_s]+i)-\widehat{X_k(i)})^2}.
\end{equation}

The multifractal analysis stems in scaling of $q$th order fluctuations so that we need to find behavior of fluctuations at scale $s$ for different values of order $q$. To do so, we construct the $q$th order fluctuation function
\begin{equation}
F_q(s)=\left(\frac{1}{2T_s}\sum_{k=1}^{2T_s}{[F^2(k,s)]^{\frac{q}{2}}} \right)^{\frac{1}{q}}.
\end{equation}
For $q=0$, the zeroth order fluctuation function is defined as
\begin{equation}
F_0(s)=\exp\left(\frac{1}{4T_s}\sum_{k=1}^{2T_s}{\log[F^2(k,s)]} \right).
\end{equation}
Order $q$ can take any real value. For $q=2$, the MF-DFA procedure reduces to DFA and it is used to analyze long-term memory properties of series $\{x_t\}$. Later in the text, we label $H\equiv H(2)$. For other values of $q$, the interpretation is not so straightforward but the scaling behavior dependence on $q$ is a basis of the multifractal analysis which we do not discuss here. In practice, minimum and maximum scales $s_{min}$ and $s_{max}$ need to be set as for finite series, the averaging and trend fitting procedures can become unreliable. Standardly, the minimum scale is set as $s_{min} \approx 10$ and the maximum scale as $s_{max}=T/4$ to avoid inefficient trend fitting for low scales and imprecise averaging at high scales.

\subsection{Useful properties of MF-DFA}

Estimation of the long-term memory parameters $H$ has a long history starting from \cite{Hurst1951}. Since then, many methods have been developed to study the power-law scaling of the autocorrelation function and the connected phenomena of the divergent at origin spectrum and the power-law scaling of variance of the partial sums. The estimators are developed in both time and frequency domains (see \cite{Taqqu1995,Taqqu1996,DiMatteo2007} for reviews of various methods).

As the MF-DFA method can be labelled as the most frequently used method of the multifractal analysis, its strengths and weaknesses have been also given an appropriate focus in the literature. None of the other methods have been studied in such detail. For our purposes, we are mainly interested in the ability of MF-DFA to deal with cycles and heavy-tailed distributions.

\cite{Hu2001} discuss the effect of trends on the properties of the detrended fluctuation analysis and a special attention is given to periodic cycles. For long-term memory processes combined with a sinusoidal trend, they show that the scaling $F_2(s)$ undergoes several cross-overs (changes in scaling rules) due to interaction between long-term memory and sinusoidal trend. For both persistent and anti-persistent series, the scaling passes through three cross-overs and the scaling laws connected to the long-term memory effects are observed for scales $s$ below the first and above the third cross-over scales. This way, it is possible to distinguish between the effect of long-term memory and sinusoidal trends. Importantly, the authors show that for the anti-persistent processes, the third cross-over scale is frequently higher than $T/4$ or even $T$ so that the long-term memory scaling needs to be obtained only from the scales below the first cross-over.

\cite{Barunik2010} study the effect of heavy tails on the most frequently used heuristic methods of the Hurst exponent estimation. They show that DFA is unbiased regardless of how heavy the tails are. For MF-DFA, they are interested in the case of $q=1$ and uncover that for reasonable tails (with tail parameter between 1.5 and 2 where the value of 2 is connected to the Gaussian distribution and the value of 1 for the Cauchy distribution), the estimates of $H(1)$ are practically unbiased as well.

In the original study, \cite{Kantelhardt2002} also discuss the possibility of highly anti-persistent processes with $H$ close to 0. In such situations, practically all estimators become severely upward-biased. However, the MF-DFA methodology is constructed for both asymptotically stationary and non-stationary processes. In practice, series $\{x_t\}$ can be integrated to a new series $\{y_t\}$ defined as $y_t=\sum_{i=1}^{t}{x_i}$ for $t=1,\ldots,T$ and MF-DFA can be applied on $\{y_t\}$. Labeling the generalized Hurst exponent of the series $\{x_t\}$ as $H_x(q)$ and the generalized Hurst exponent of the integrated series $\{y_t\}$ as $H_y(q)$, it holds that $H_x(q)=H_y(q)-1$. Therefore, if $\{x_t\}$ possesses properties resembling strong anti-persistence, the generalized Hurst exponent for the series can be obtained from running MF-DFA on the integrated series and reducing the estimate by 1.

DFA as a special case of MF-DFA is thus an ideal candidate for the long-term memory analysis of the electricity prices as the above mentioned properties match with the properties of the electricity prices discussed in the previous sections as well as in the next section dealing with specific properties of the Czech electricity spot prices.

\begin{table}[htbp]
\centering
\caption{Descriptive statistics}
\label{tab1}
\footnotesize
\begin{tabular}{c|c|c}
\toprule \toprule
&price&change in price\\
\midrule \midrule
mean&43.7700&0.0001\\
SD&16.2551&6.6093\\
skewness&0.1040&0.4136\\
excess kurtosis&1.6686&7.4503\\
\midrule
Shapiro-Wilk test&14.9120&19.6280\\
$p$-value&$<0.01$&$<0.01$\\
Jarque-Bera test&4041&80315\\
$p$-value&$<0.01$&$<0.01$\\
\midrule
ADF - test (50)&-13.9594&-36.1585\\
$p$-value&$<0.01$&$<0.01$\\
KPSS (50)&8.7456&0.0012\\
$p$-value&$<0.01$&$>0.1$\\
\bottomrule
\end{tabular}
\end{table}

\section{Data description}

\begin{figure}[htbp]
\center
\begin{tabular}{cc}
\includegraphics[width=3in]{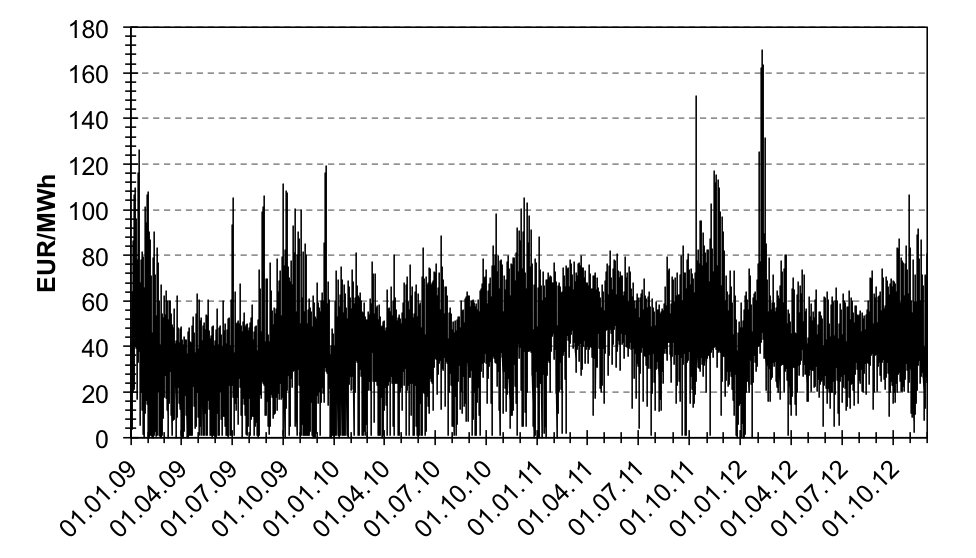}&\includegraphics[width=3in]{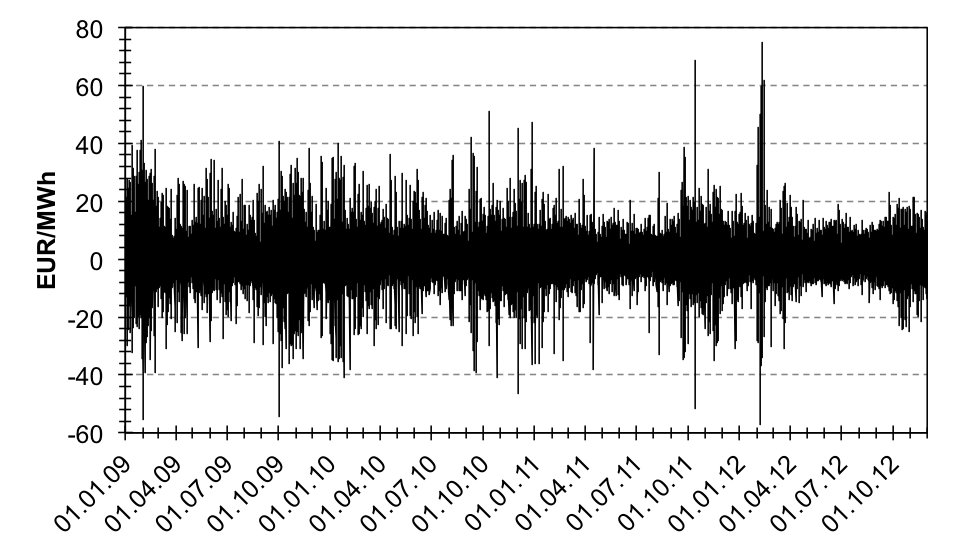}\\
\end{tabular}
\caption{\footnotesize\textbf{Time series plots.} Electricity hourly prices (left) and their changes (right) are shown. The changes resemble returns of various financial assets whereas dynamics of the prices is much further from these. \label{ts}}
\end{figure}

\begin{figure}[htbp]
\center
\begin{tabular}{cc}
\includegraphics[width=3in]{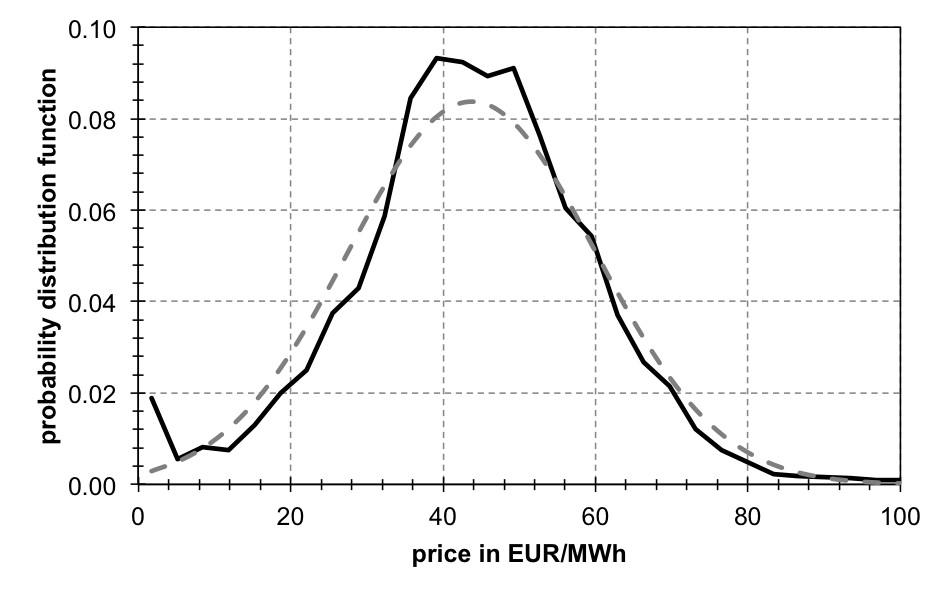}&\includegraphics[width=3in]{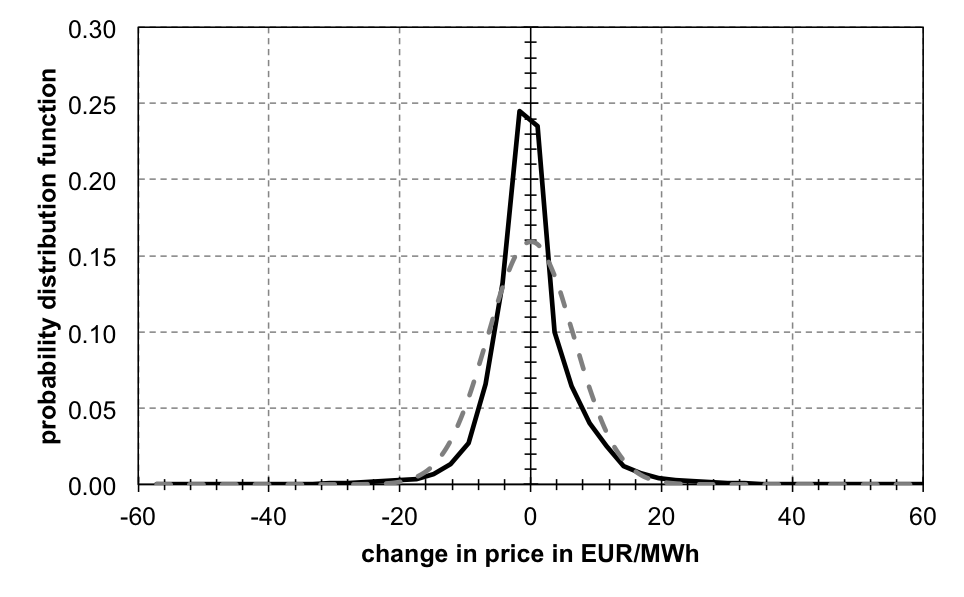}\\
\end{tabular}
\caption{\footnotesize\textbf{Histograms of electricity prices and changes in prices.} Probability distribution function of prices (left, in black) is quite close to the normal distribution (dashed grey line) with the exception of the censored left tail (no negative prices in the sample). Changes in prices (right) are much further from normality.\label{hist}}
\end{figure}

We analyze hourly electricity spot prices in the Czech Republic between 2009 and 2012, namely 1.1.2009 - 30.11.2012, with a total of 34,316 observations\footnote{Data were obtained from Yearly Reports of OTE -- the electricity market operator in the Czech Republic -- available at http://www.ote-cr.cz/statistics.}. The prices are denominated in EUR per MWh and negative prices were not allowed before 2013. In Fig. \ref{ts}, we show evolution of prices during the analyzed period. It is evident that prices jump frequently in both directions. The first differences of the prices strongly resemble standard returns of stocks or exchange rates with volatility clustering and extreme movements. The first differences are far from being normally distributed as shown in Fig. \ref{hist}. However, the original price series are close to normally distributed if we omit the fact that the prices are censored from below. Overall non-normality of distributions is supported by Shapiro-Wilk \citep{Shapiro1965} and Jarque-Bera \citep{Jarque1980,Jarque1981} tests in Table \ref{tab1} which shows strong rejection of normality for both series. Standard descriptive statistics support only mild heavy tails of prices but heavy tails for the first differences. Both series are positively skewed so that the more extreme upward movements are more likely. However, the skewness of prices is very close to zero hinting symmetry which is again in hand with the histograms in Fig. \ref{hist}.

For the analysis of dynamics of the series, distinguishing between stationary and non-stationary series is crucial. To this point, we utilize the ADF \citep{Dickey1979} and KPSS \citep{Kwiatkowski1992} tests. The null hypothesis of the former is a unit root against no unit root whereas for the latter, the hypothesis of stationarity against non-stationarity is tested which provides an ideal combination of tests. Results presented in Table \ref{tab1} give an evidence that the price series are non-stationary but do not contain a unit root whereas the first difference series are stationary. In terms of the long-term memory notation, the prices of electricity are in the interval $1<H<1.5$ and thus the first differences in $0<H<0.5$. We thus follow with analysis of prices and not the first differences due to two reasons. First, we do not want to loose information about the dynamics of the prices which would be done by first differencing. And second, the price series are much more interesting in the electricity context, there are no actual returns to the series as it is not possible to buy a MWh of electricity and sell it the following period.

\begin{figure}[htbp]
\center
\begin{tabular}{cc}
\includegraphics[width=3in]{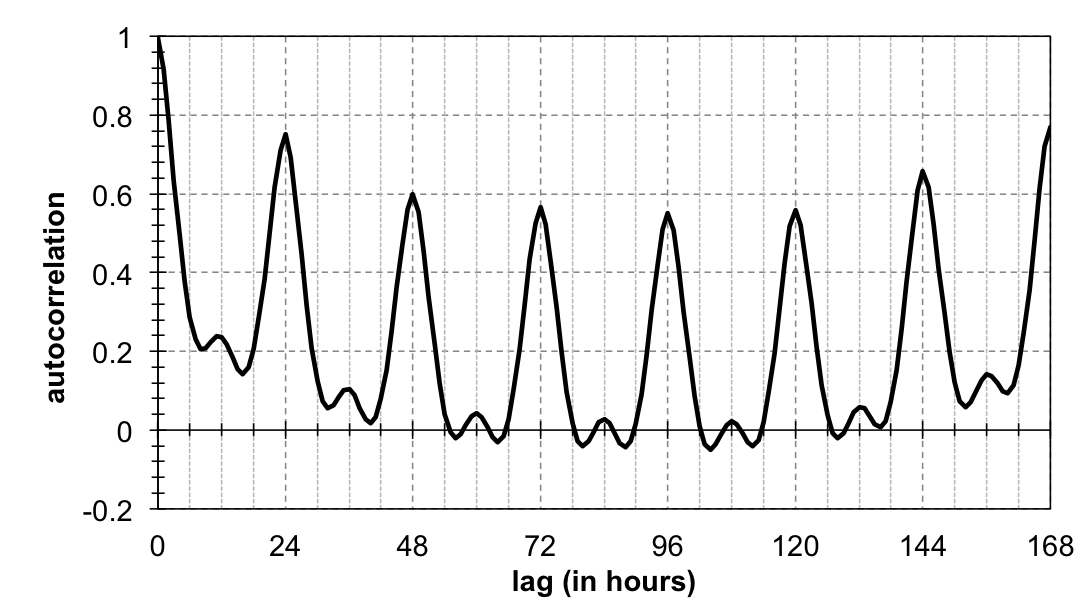}&\includegraphics[width=3in]{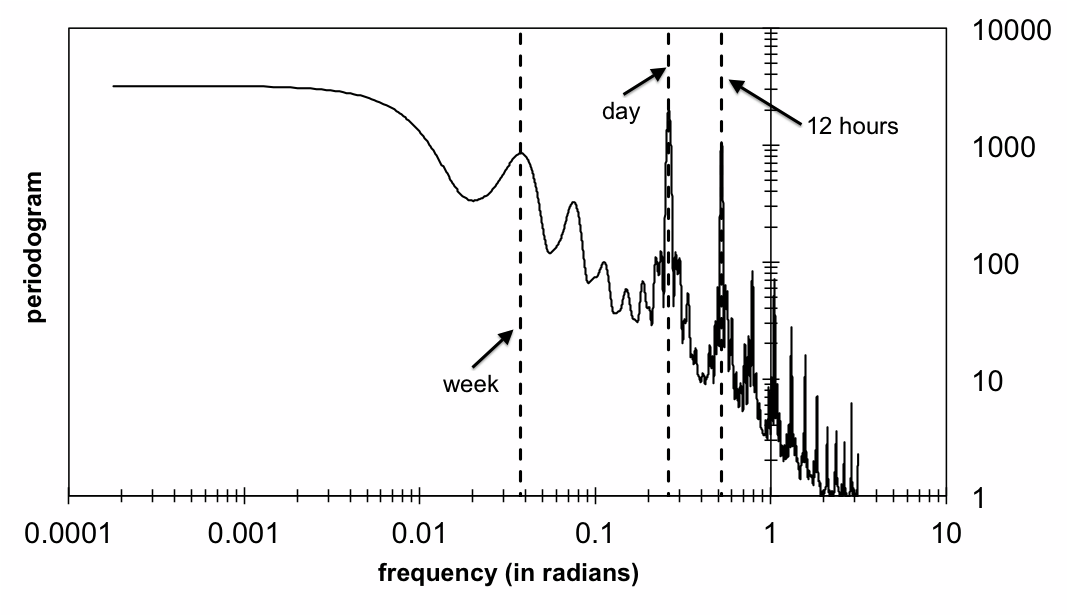}\\
\end{tabular}
\caption{\footnotesize\textbf{Correlation structure of prices.} Both autocorrelation function (left) and periodogram (right) show strong seasonal component with a dominating scale of 24 hours. \label{ACFperiod}}
\end{figure}

\begin{figure}[htbp]
\center
\begin{tabular}{cc}
\includegraphics[width=3in]{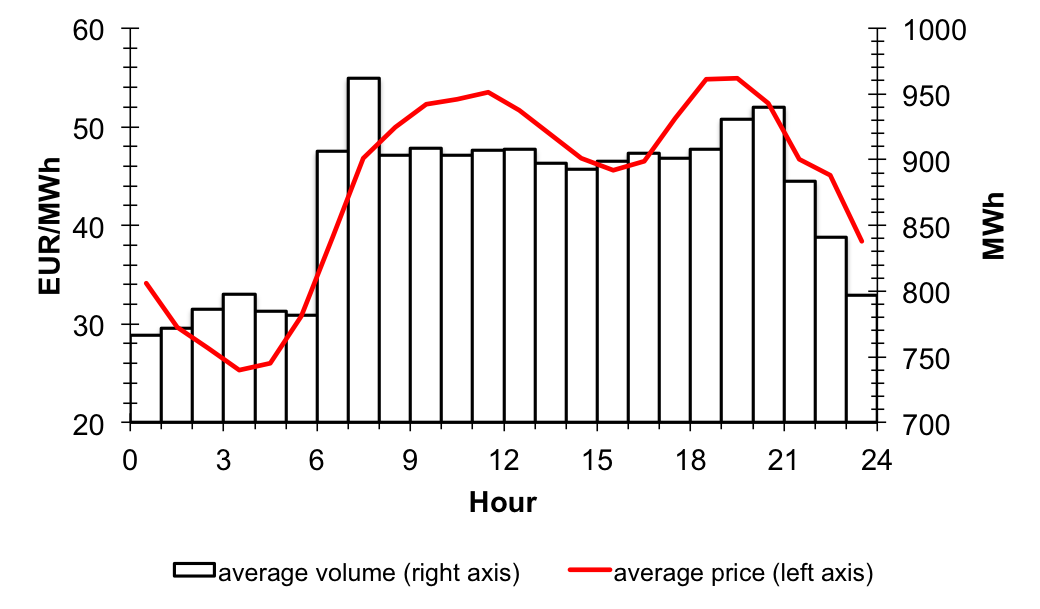}&\includegraphics[width=3in]{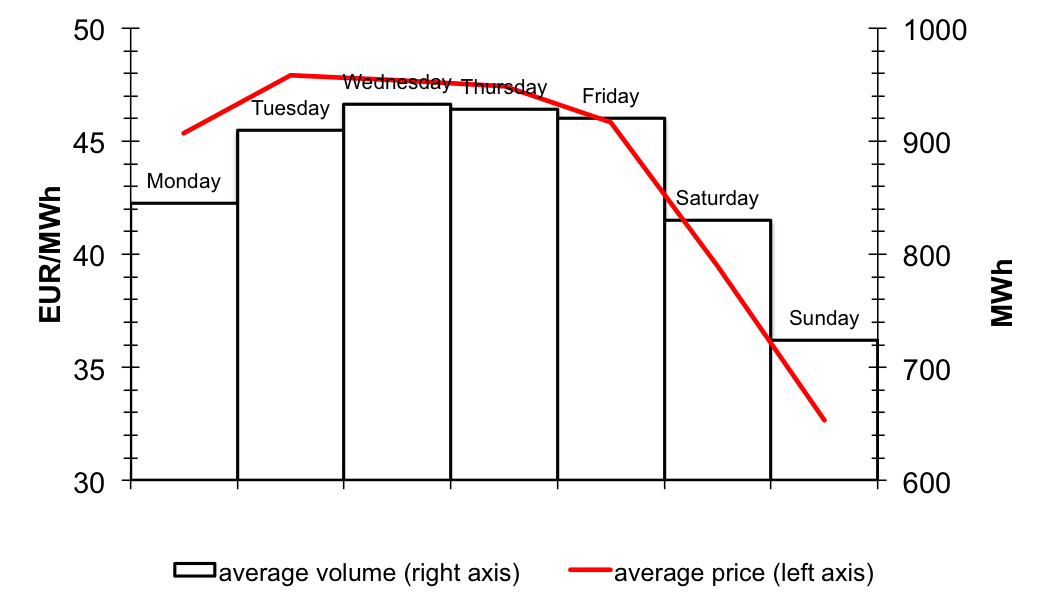}\\
\includegraphics[width=3in]{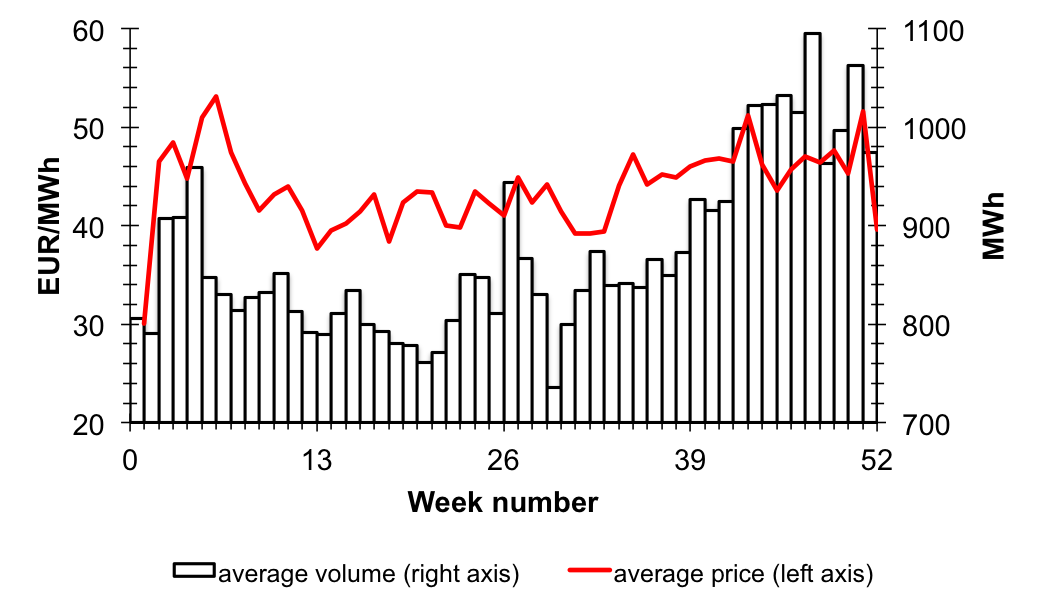}&\includegraphics[width=3in]{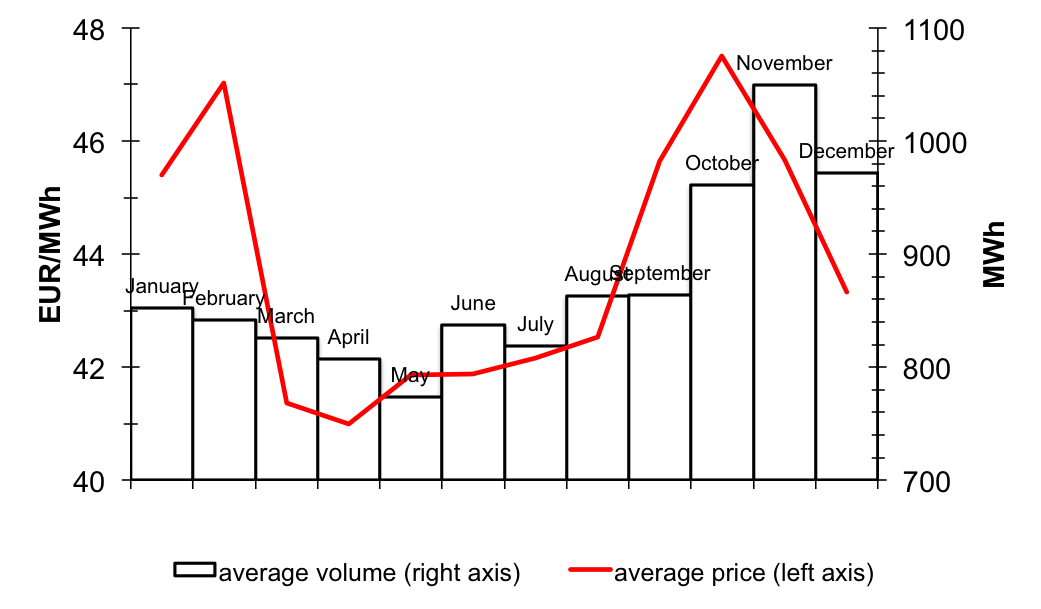}\\
\end{tabular}
\caption{\footnotesize\textbf{Cyclical properties of electricity prices and volumes.} Seasonal patterns are shown for intraday (upper left), daily (upper right), weekly (lower left) and monthly (lower right) scales. Apart from the weekly scale, both prices and volumes show pronounced seasonal patterns. \label{cycles}}
\end{figure}

The memory properties of the electricity prices are further illustrated by sample auto-correlation function and periodogram\footnote{Periodogram is based on Bartlett weights with a bandwidth of 370, i.e. approximately 0.1 of the time series length.} in Fig. \ref{ACFperiod}. There, we observe that the dynamics of the prices is very cyclical with a dominating frequency of 24 hours. Both the auto-correlation function and periodogram are well in hand with the definitions of long-term memory. However, we can see that both the power-law decay of the auto-correlation function and the power-law divergence at the origin of the periodogram are disturbed by the aforementioned cyclical properties. Due to this fact, we utilize DFA to analyze the long-term memory properties of the series as discussed in the previous section. The complex cyclicality is further illustrated in Fig. \ref{cycles} where we show how the average price and average traded volume depend on an hour of the day, day of the week, and week and month of the year. This again calls for a robust method of the Hurst exponent estimation as discussed previously.

\section{Results and discussion}

\begin{figure}[htbp]
\center
\begin{tabular}{c}
\includegraphics[width=5in]{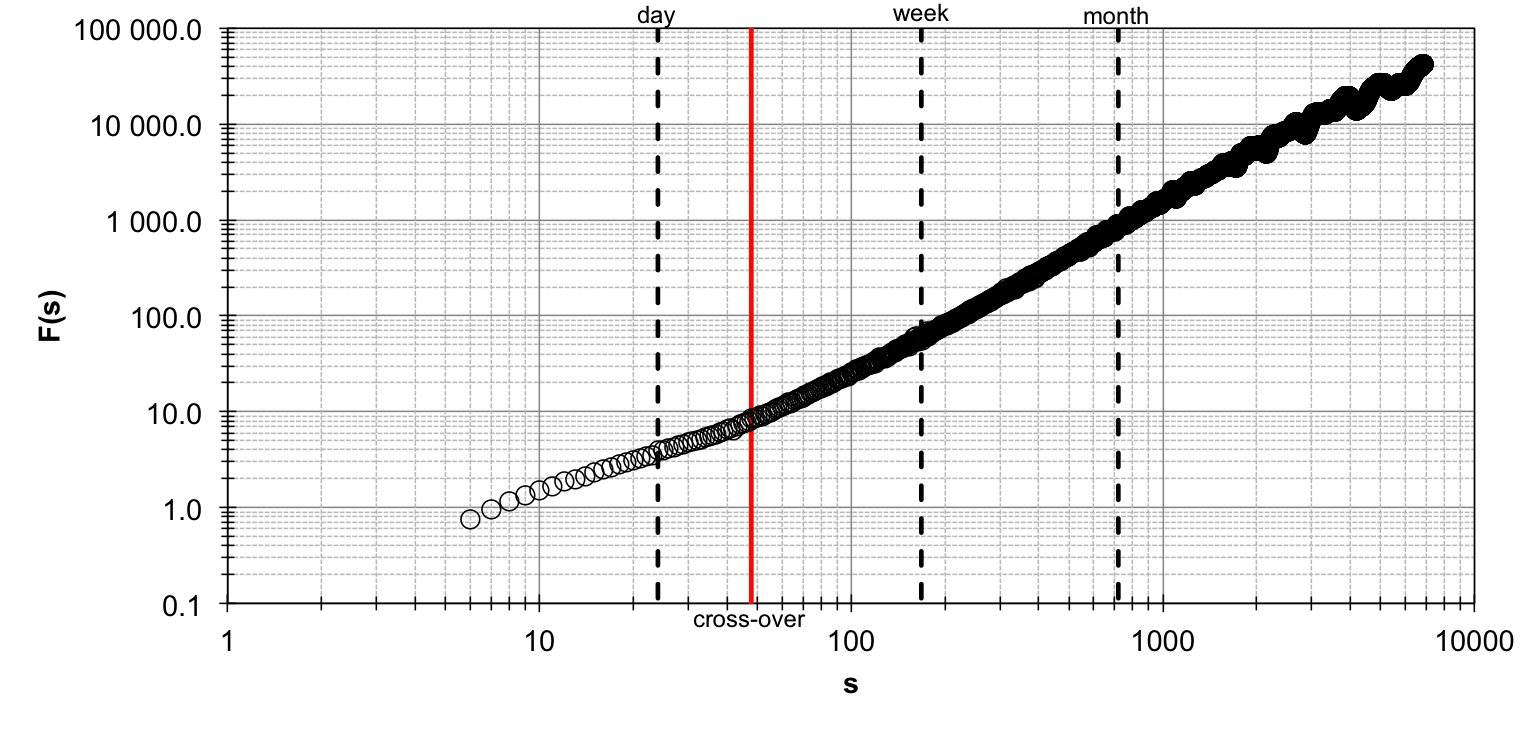}\\
\end{tabular}
\begin{tabular}{cc}
\includegraphics[width=3in]{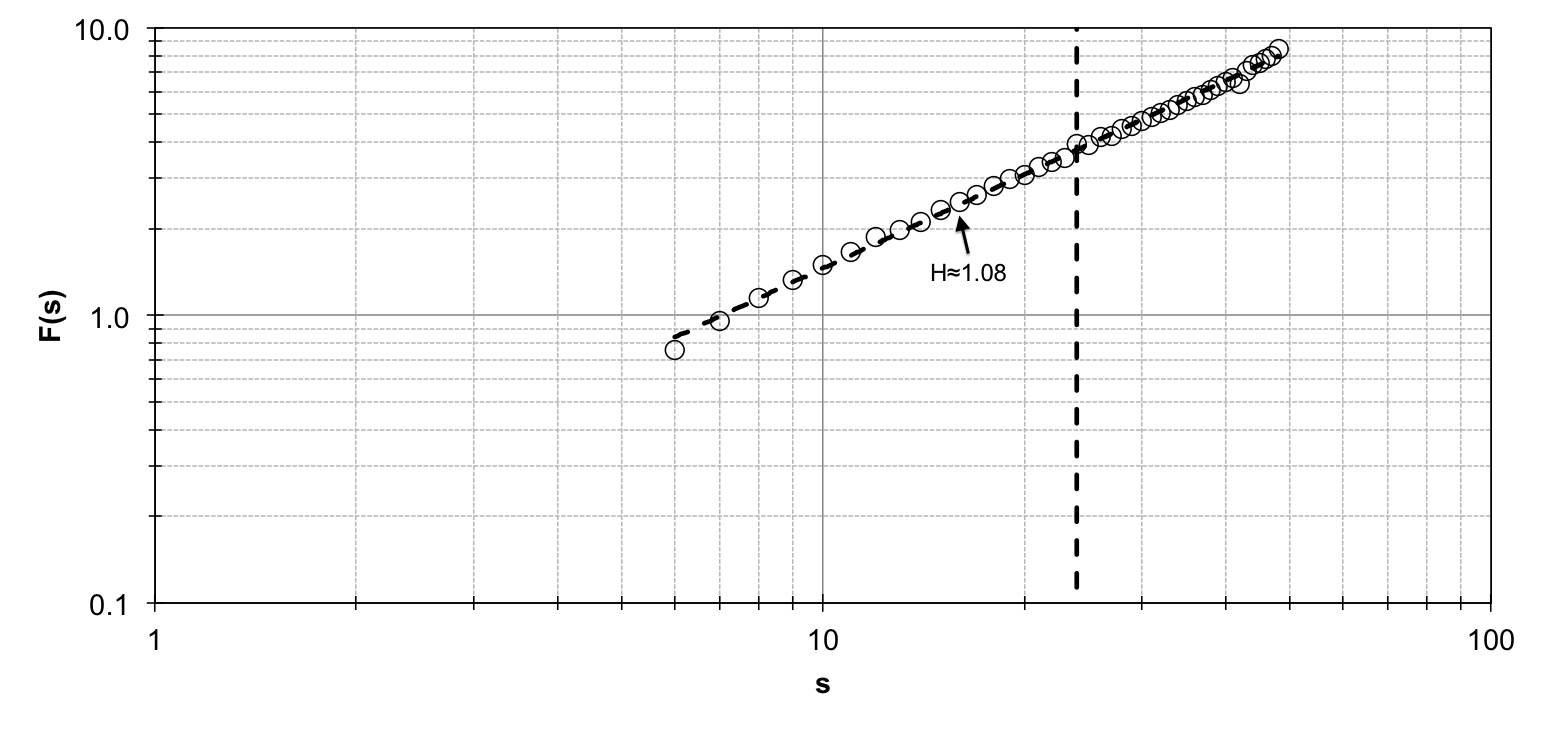}&\includegraphics[width=3in]{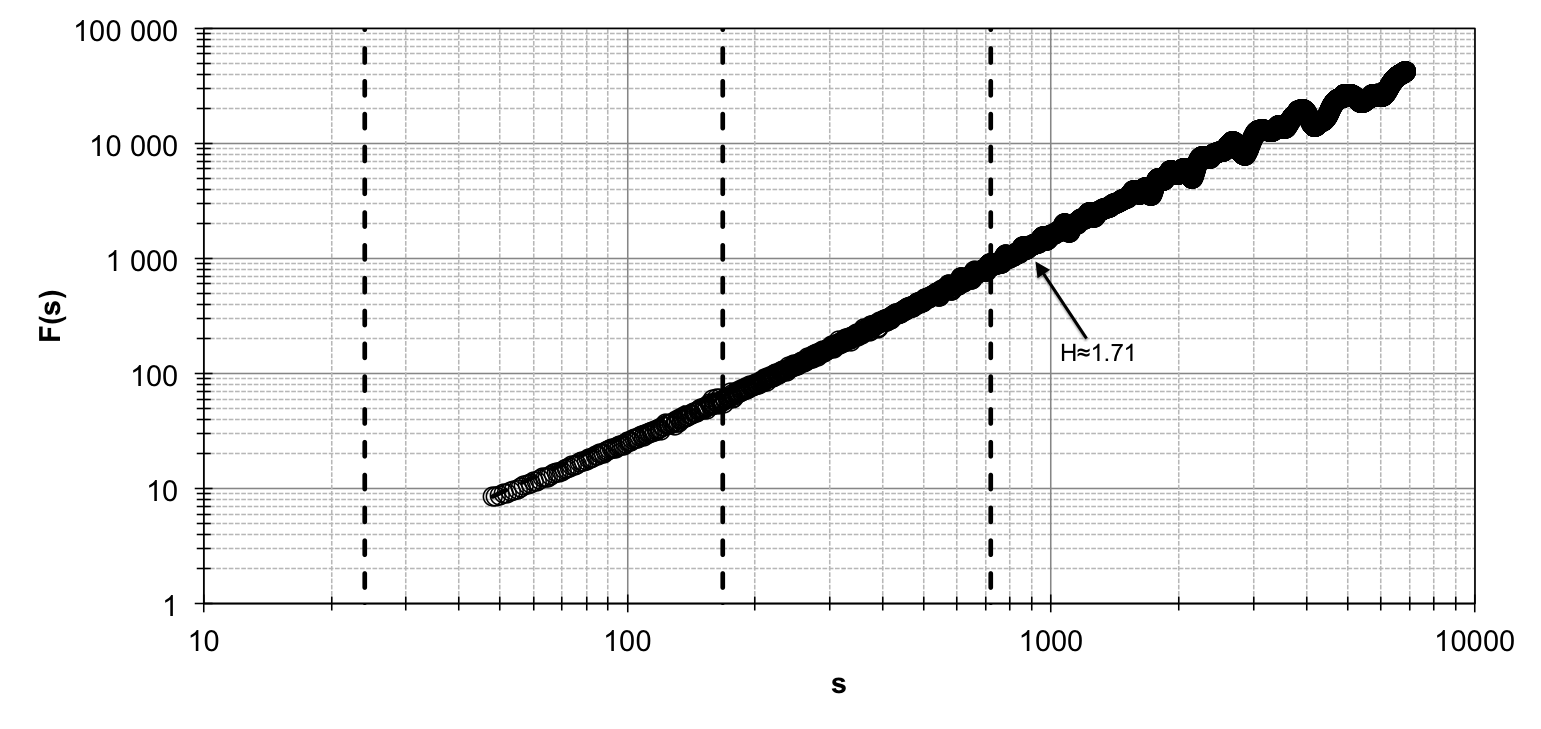}\\
\end{tabular}
\caption{\footnotesize\textbf{Scaling of fluctuations $F(s)$.} The upper panel shows the scaling of the fluctuation function and a pronounced cross-over at approximately two days. The lower panels show the scaling for both regimes in more detail. The lower left panel characterized by $H=1.08$ is attributed to the long-term memory of the prices process and the lower right panel shows scaling for the scales dominated by cyclical components. \label{F}}
\end{figure}

We analyze long-term memory in electricity spot prices of the Czech system using detrended fluctuation analysis, i.e. multifractal detrended fluctuation analysis with $q=2$. Setting $s_{min}=6$ and $s_{max}=T/4$ for the MF-DFA, we obtain scaling of the fluctuation $F_2(s)$ illustrated in Fig. \ref{F}. As the data frequency equals one hour, the minimum scale is set at a quarter of a day and the maximum scale approximately matches a year. Based on the initial analysis of the series in the Data description section, we assume that the series contain strong cycles but might also possess long-term memory. It is thus reasonable to assume that the scaling of $F_2(s)$ contains at least one cross-over. This is indeed true for the analyzed electricity prices as shown in Fig. \ref{F}. We observe one evident cross-over at $s_{\times}\approx 48$. The cross-over splits the scaling chart into two laws which resemble a power-law scaling strongly as shown in the split charts in Fig. \ref{F}. This gives two Hurst exponents -- $H\approx 1.1$ for $s\le48$ and $H\approx 1.7$ for $s\ge48$. Note that these Hurst exponents do not differ considerably for varying $s_{\times}$ between 36 (1.5 day) and 72 (3 days) and these are thus quite stable. This multi-scaling can be attributed to competing effects of the long-term memory and periodic trends which are both strong parts of dynamics of the electricity prices. As discussed in the previous section, the Hurst exponent based on scales below the first cross-over scale $s_{\times}$ can be used for interpretation of the long-term memory. Therefore, the price dynamics is characterized by $H\approx 1.1$ and the prices are thus strongly persistent and non-stationary but still remain well below the unit-root level of $H=1.5$ so that they remain mean-reverting. This is well in hand with the basic description in Tab. \ref{tab1}.

The persistence of the series implies that prices follow rather long-lasting trends, which are even above standard long-term memory with $0<H<1$ making the prices non-stationary. Nonetheless, the dynamics is far from the unit-root behavior and the prices return to their long-term levels. Such behavior is very different from other financial assets which usually follow a random walk and their returns are thus unpredictable (or at least not systematically predictable). However, we need to keep in mind that such a persistence of electricity prices cannot be easily exploited for profit. The persistence can be also seen as a product of incorrect expectations about a future need of electricity of the market agents. Remembering that the electricity spot market exists to cover the unexpected demand for electricity (as the majority of supplied electricity is based on medium- and long-term contracts), the extreme price movements are mainly caused by external unexpected events (temperature, humidity, macroeconomic news, etc.). When the unexpected event comes, it usually has a medium- or long-lasting effect (e.g. temperature above long-term averages is usually characteristic for whole day or even longer periods) but the agents cannot ``pre-buy'' the electricity quickly. To cover the increased demand, additional (and usually more expensive) power sources need to be connected to the network which increases the electricity prices. The combined effect of non-storability and connecting less efficient power sources pushes the electricity prices to the persistent behavior.

\begin{figure}[htbp]
\center
\begin{tabular}{c}
\includegraphics[width=5in]{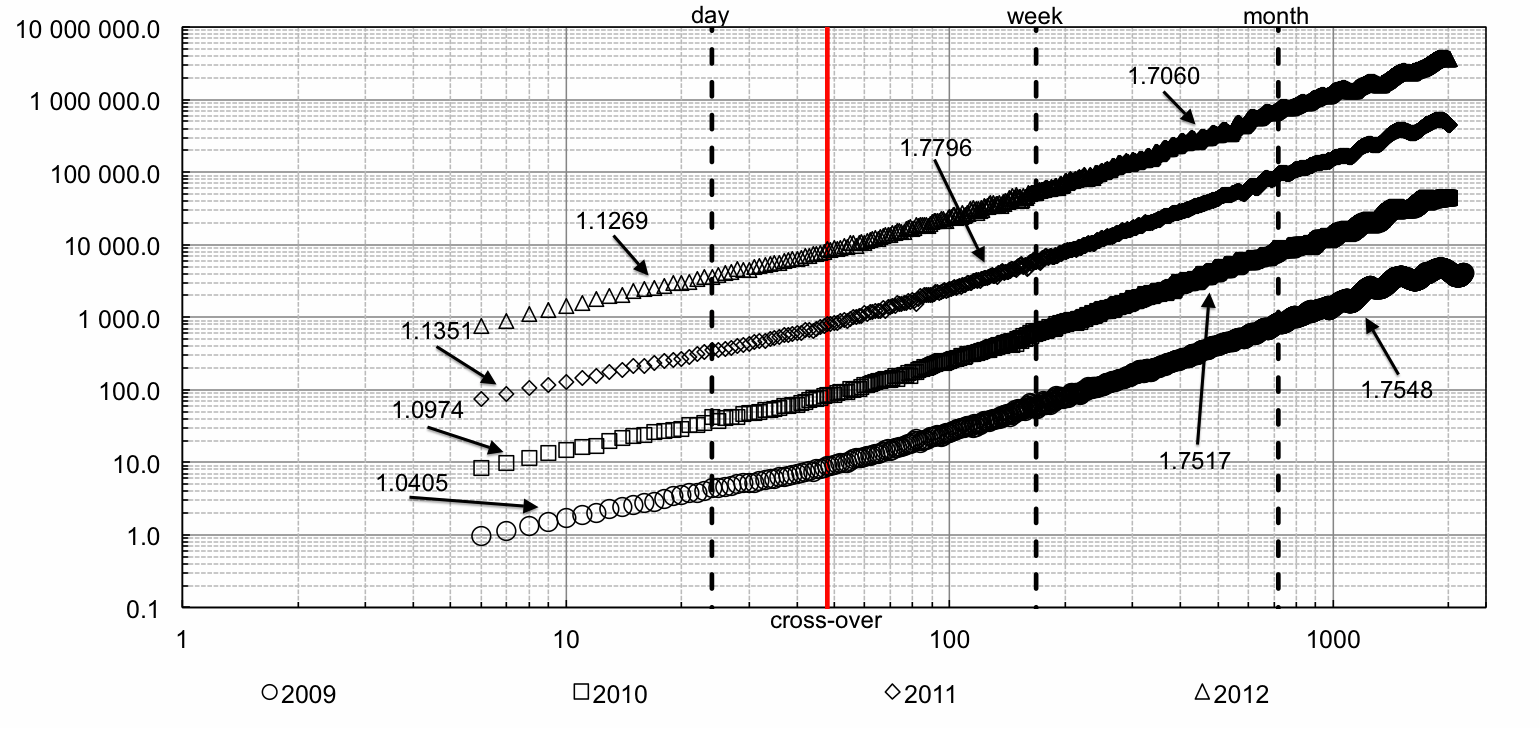}\\
\end{tabular}
\caption{\footnotesize\textbf{Scaling of fluctuations $F(s)$ for separate years.} Estimated Hurst exponent are shown for two scaling regimes for separate years between 2009 and 2012. Scaling exponents are remarkably stable. \label{F_years}}
\end{figure}

To see whether these properties are stable in time, we analyze the long-term memory components in the same way but for the separate years 2009-2012. In Fig. \ref{F_years}, we observe that the Hurst exponent connected to the long-term memory is rather stable and approximately around 1.1 for the price series. For the higher scales, we again see stability of the scaling exponent around 1.75. The non-stationary mean-reverting persistence is thus observed even for separate years. Note that only the very specific characteristics of the DFA method allow us to study the long-term memory without arriving at spurious results. Standardly, the Hurst exponent connected to the higher scales would be reported. However, the difference between having $H<1.5$ and $H>1.5$ is crucial. For the former, the prices return to their long-term mean. But for the latter, the prices would explode. Note that having $H \approx 1.1$ implies that the mean reversion is very rapid. These characteristics are very stable in time.

\section{Conclusion}

We have analyzed long-term memory properties of hourly spot prices of the Czech electricity between 2009 and 2012. As the electricity prices have very intriguing properties, such analysis is rather challenging. We have shown that the Czech prices follow similar patterns observed for other electricity prices, mainly intraday, daily and monthly patterns in both prices and volume. Utilizing the detrended fluctuation analysis, we have been able to separate these cyclical properties from the long-term memory. The results are in hand with majority of the relevant literature as we show that the electricity prices are non-stationary but mean-reverting so that their behavior is partly predictable. However, due to specific features of electricity (mainly its non-storability), such a predictable behavior cannot be easily exploited for earning profits. The electricity prices are thus very different from standard financial assets such as stocks or exchange rates and they need to be treated accordingly. The found patterns of behavior of the electricity prices can be attributed to their structure as the analyzed series contained the spot prices. These serve mainly to balance demand for electricity which has not been covered by futures contracts. As such, the unexpected change in demand for electricity is rather short-lived and the reversion to a long-term price is quite rapid which is represented by the Hurst exponent close to (but higher than) unity. Stability of the results in specific years and correspondence to the results of more developed markets underline that the Czech electricity market has reached a similar levels of development.

\section*{Acknowledgments}
 
The support from the Czech Science Foundation under Grants 402/09/0965 and P402/11/0948, and project SVV 267 504 is gratefully acknowledged.
 
%\section*{References}
\bibliography{Electricity}
\bibliographystyle{chicago}

\end{document}